\documentclass[aps,prl,twocolumn,groupedaddress]{revtex4}
\usepackage{graphicx}

\begin{document}

\title{Elastic scattering losses in the four-wave mixing of Bose Einstein
Condensates}

\author{J. Chwede\'{n}czuk$^1$, M. Trippenbach$^1$ and K. Rz\c{a}\.zewski$^2$}

\affiliation{$^1$ Physics Department, Warsaw University, Ho\.{z}a 69, PL-00-681 Warsaw,
Poland, $^2$ Center for Theoretical Physics, Polish Academy of Science, Al. Lotnik\'{o}w,
PL-00-681 Warsaw, Poland.}

\begin{abstract}
We introduce a classical stochastic field method that accounts for the quantum
fluctuations responsible for spontaneous initiation of various atom optics processes. We
assume a delta-correlated Gaussian noise in all initially empty modes of atomic field.
Its strength is determined by comparison with the analytical results for two colliding
condensates in the low loss limit. Our method is applied to the atomic four wave mixing
experiment performed at MIT [Vogels {\it et. al.}, Phys. Rev. Lett. {\bf 89}, 020401,
(2002)], for the first time reproducing experimental data.
\end{abstract}

\maketitle

PACS numbers 03.75.Hh, 05.30.Jp

\bigskip

In recent years we observe a growing number of experiments in which the atomic
Bose-Einstein condensate evolves in a nontrivial way. A whole new area of nonlinear atom
optics was born. The most striking example of such a nonlinear process is the atomic
four-wave mixing (4WM). In close analogy to its optical counterpart, atomic 4WM consists
of generation of the new atomic beam in the nonlinear interaction of three overlapping
matter waves. For the main part atomic 4WM is an example of a stimulated process.
However, during this process there are also collisions between individual atoms that lead
to a population of initially unoccupied atomic states. These processes have spontaneous
initiation but by nature they are also examples of the four particle mixing. They amount
to the elastic scattering losses from the coherently  evolving condensates.

The standard tool used to describe dynamics of the condensate within mean field
approximation is the celebrated Gross-Pitaevskii equation (GPE). As it stands, it is
capable of describing stimulated processes but not the spontaneous ones. However, at
least in some experiments ~\cite{Vogels} the elastic scattering losses may become
significant. There are at least two theoretical attempts to estimate such losses during
the collision of condensates. In the first one ~\cite{Band} the authors used
momentum-dependent higher order correction to the nonlinear coupling constant in the GPE,
introducing complex scattering length. In the second one ~\cite{Yuro,Keith,Bach} the
field theoretical formulation was used. To make it effective, the authors approximate the
second quantized hamiltonian by a quadratic form. Both methods give very similar results
but are applicable only if the elastic scattering losses are merely a small correction.

It is the purpose of this Letter to formulate a general method of describing elastic
scattering losses in the nonlinear atom optics processes. To this end we add to the GPE a
classical gaussian noise, representing vacuum quantum fluctuations of the atomic field
and an auxiliary field holding atoms scattered out from BEC. Such a technique has its
roots in quantum optics.

Spontaneous optical processes have their origins in quantum fluctuations. The best known
example is a process of superfluorescence ~\cite{Vrehen}. In this phenomenon a sample of
atoms is prepared in the internal excited state. Spontaneously emitted photons create an
avalanche of photon emissions. When the light field becomes strong, it is well described
by a classical electromagnetic field. However the initiation has a quantal nature. This
quantum initiation was successfully imitated by a classical noise ~\cite{Haake}. There
are also general methods of mapping quantum fluctuations into stochastic term in the
evolution equations of quantum optics (generalised P-representation methods)
\cite{drumm}.

In optics one can find numerous other processes initiated by spontaneous emission and
eventually upon populating empty modes turning into stimulated processes; eg. spontaneous
Raman scattering \cite{Raman}, parametric down conversion \cite{paradown}, etc. There are
also similar examples in atomic and molecular physics \cite{molec}. Our method is general
and is capable of treating many of these processes. Here we demonstrate the method using
the 4WM of coherent matter waves.

The first experiment demonstrating 4WM in a sodium Bose-Einstein condensate was performed
at NIST~\cite{Deng}. This was followed by a theoretical and numerical treatment of the
experiment ~\cite{Trip,Gold}. In the experiment, a short time of free expansion of the
condensate, after it was released from the magnetic trap, was followed by a set of two
Bragg pulses \cite{Kozuma}, which created moving wavepackets. These wavepackets, together
with the remaining stationary condensate, due to nonlinear interaction and under phase
matching conditions created a new momentum component in the 4WM process. The standard
starting point for the description of atomic 4WM process is the Gross-Pitaevskii equation
\begin{eqnarray}\label{GP}
i\hbar
\partial_t \Psi(\vec r, t)&=&\left(-\frac{\hbar^2 \nabla^2}{2m}+V+gN|\Psi|^2 \right)
\Psi(\vec r, t).
\end{eqnarray}
Here $N$ is the total number of atoms, $|\Psi|^2$ is proportional to the atomic number
density and is normalized to one, $g=4\pi\hbar^2 a/m$ is the nonlinear interaction
strength, $m$ the atomic mass, $a$ is the scattering length and $V$ is a confining
potential. A compact ground state wavefunction $\Psi(\vec r, 0)$ is created in harmonic
trap potential $V$ and centered around $ r= 0$ with $\Psi(0,0)=\Psi_m$, the maximum
value. Once this ground state is created, V is turned off. The development of $\Psi(\vec
r, t)$ is now described by Eq.~(\ref{GP}) with $V=0$. Later on, a set of Bragg pulses is
applied and parts of the condensate begin to move. We can define two timescales
characterizing evolution of the condensate: a nonlinear interaction time $\tau_{{\rm
NL}}=(gN|\Psi_m|^2/\hbar)^{-1}$ and a collision time $\tau_{{\rm col}}$. The latter is
defined as a time it takes two wavepackets uniformly moving along $x$ to move apart (so
they just touch and cease to overlap), $\tau_{{\rm col}}=2r_{{\rm TF}}/v$ where $r_{{\rm
TF}}$ is the initial radius of the condensate in the $x$ direction (Thomas - Fermi
approximation), and $v$ is the relative velocity. The ratio of these two timescales
determines the output of the 4WM process.

The initial condition immediately after application of the Bragg
pulses at $t_1$, can be approximated as being a composition of the
BEC wavepackets, identical in shape to $\Psi(\vec r, t_1)$ (for
more details see \cite{Trip}):
\begin{eqnarray}
\Phi(\vec r, t_1)&=&\Psi(\vec r, t_1)\sum_{i=1}^3f_i^{1/2} {\rm
e}^{i\vec P_i\vec r/\hbar}.
\end{eqnarray}
Here $f_i=N_i/N$ is the fraction of atoms in the $i$-th wavepacket and
$\sum_{i=1}^3f_i=1$. A new wavepacket with $\vec P_4=\vec P_1-\vec P_2+\vec P_3$ will
build up, thanks to the nonlinear interactions accounted for by the last term in the
Gross Pitaevskii equation (\ref{GP}). After a while, the fourth wavepacket will grow to
the macroscopic level. Using the de Broglie relations: $\vec k_i=\vec P_i/\hbar$ and
$\omega_i=\hbar k_i^2/2m$ we have:
\begin{eqnarray}
\label{PSI}\Phi(\vec r, t)&=&\sum_{i=1}^4\Phi_i(\vec r, t) {\rm
e}^{i(\vec k_i\vec r-\omega_it)},
\end{eqnarray}
with initial conditions
\begin{eqnarray}
\Phi_i(\vec r, t_1)=f_i^{1/2}\Psi(\vec r, t_1),\,\,\, i=1,2,3; && \Phi_4(\vec r, t_1)=0.
\nonumber
\end{eqnarray}
Variation of the $\Phi_i$ is assumed to be slow as compared to that of the exponential in
equation (\ref{PSI}). Four equations for this slow dependence are obtained from
(\ref{GP}) and (\ref{PSI}). We have, when $V$ is turned off \cite{Band}:
\begin{eqnarray}
\imath\hbar\partial_t\psi_{i}&=&-\frac{\hbar^2}{2m}\nabla^2\psi_{i}
+gN(|\psi_{i}|^2+2\sum_{j\neq
i}|\psi_{j}|^2)\psi_{i}\nonumber\\
&+&2gN\psi_{i+1}\psi_{i+2}^*\psi_{i+3} \label{cztery}
\end{eqnarray}
where we use the convention in which all indices are taken modulo 4. To account for the
elastically scattered atoms we introduce an additional component of the wavefunction
$\psi_B$. It is this part of the wavefunction which will be initiated by the classical
stochastic field. The stochastic field $\xi_{ij}(\vec r,t)$ must be added to the equation
of motion (\ref{cztery}) to trigger the elastic scattering process of two particles from
the condensate wavefunctions $\psi_i$ and $\psi_j$ to the background field $\psi_B$. It
is a four particle process and it must be implemented for each pair of colliding
wavefunctions $i,j$ in such a way that the total number of atoms in colliding waves +
background field is still conserved. The resulting set of equations reads
\begin{eqnarray}
\label{4wm_1}(\imath\hbar\partial_t+\frac{\hbar^2}{2m}\nabla^2)\psi_{i}&=&gN
(|\psi_{i}|^2+2\sum_{j\neq i}|\psi_{j}|^2+2|\psi_B|^2)\psi_{i}\nonumber\\
&+&gN\sum_{j\neq i}\psi_{j}^*\psi_B(\psi_B+\xi_{ij})\nonumber\\
&+&2gN\psi_{i+1}\psi_{i+3}\psi_{i+2}^*
\end{eqnarray}
\begin{eqnarray}
\label{4wm_2}(\imath\hbar\partial_t+\frac{\hbar^2}{2m}\nabla^2)\psi_B&=&
+gN(|\psi_B|^2+2\sum_{i=1}^4|\psi_i|^2)\psi_{B}\nonumber\\
&+&gN \sum_{i \neq j}\psi_{i}\psi_{j} (\psi_B+\xi_{ij})^*
\end{eqnarray}
For numerical calculations we assume that $\xi_{ij}(\vec r,t)$ is a gaussian stochastic
process with zero mean and the only nonvanishing second order correlation function equal
to $\langle \xi_{ij}^*(\vec r,t)\xi_{ij}(\vec r',t')\rangle=A_{ij}\delta_{\vec r,\vec
r'}\delta_{t,t'}$. Here Kronecker delta functions are assumed both in space and time
since we refer to numerical simulations with spacial grid and discrete time steps. Notice
that we assign different stochastic process to each pair of colliding wavepackets,
anticipating dependence on parameters like relative velocity.

\begin{figure}
  \centering
  \includegraphics[scale=0.35]{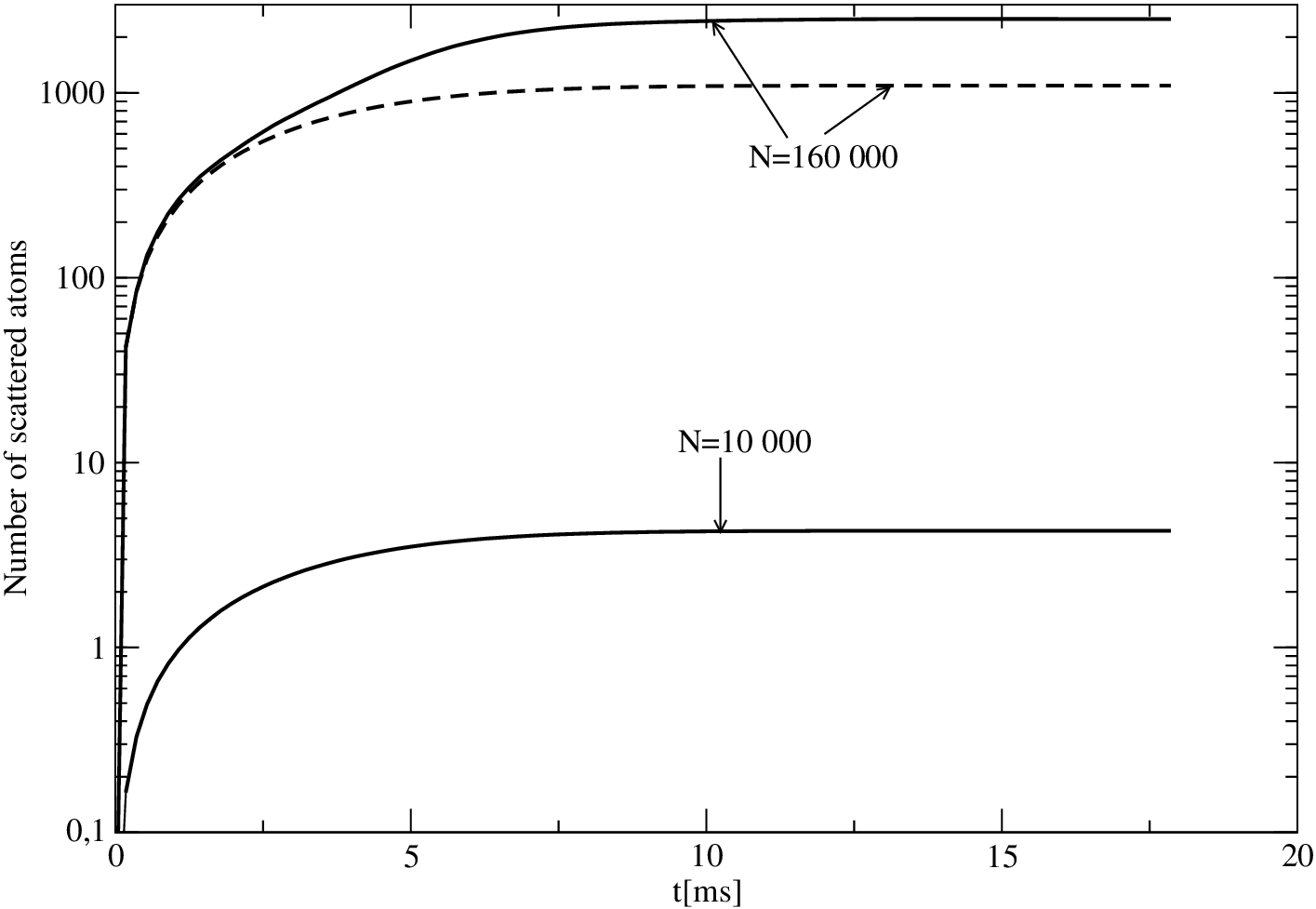}
  \caption{Number of elastically scattered atoms
from the pair of counter-propagating condensate Gaussian wavefunctions, with relative
velocity $1.75\,mm/s$, as a function of time. The lower curve corresponds to small number
of scattered atoms ($N=10^4$ and gaussian width $\sigma=9.1\,\mu m$), when the bosonic
stimulation does not occur. In this limit all three methods: complex scattering length
calculation \cite{Band}, field theory \cite{Bach}, and our stochastic method give
indistinguishable results. The upper pair of curves corresponds to higher number of atoms
($N=1.6\cdot10^5$ and gaussian width $\sigma=15.8\,\mu m$) - dashed line was obtained
using complex scattering length method and the solid line is a solution of
eq.(\ref{psi_b})}
  \label{fig1}
\end{figure}

Equations (\ref{4wm_1}-\ref{4wm_2}) may be obtained from multiatom system hamiltonian
upon using Bogoliubov decomposition of the atomic field operators into condensate parts
and initially empty modes $\psi=\sum_{i=1}^4\psi_i+\psi_B$ in a way analogous to that
presented in \cite{Bach}. We do not however explicitly decompose $\psi_B$ into plane
waves but obtain it's Heisenberg equation of motion assuming only that it commutes with
operators of macroscopically occupied modes $\psi_i$. Finally stochastic field is added
to $\psi_B$ in the source terms for 4WM as shown above. These stochastic terms mimic
vacuum quantum fluctuations leading to spontaneous elastic scattering loss. But as
elastically scattered atoms reside in $\psi_B$ they may eventually get amplified via
bosonic stimulation when population becomes significant. This is an outline of our
stochastic method; details will be presented elsewhere \cite{Jachwed}.

To fully determine equations (\ref{4wm_1}-\ref{4wm_2}) we need to specify the value of
constants $A_{ij}$. Just as it has been done in the case of superfluorescence mentioned
above, we can find $A_{ij}$ by the requirement that it reproduces known limiting analytic
results ~\cite{Bach}. In reference ~\cite{Bach} the elastic scattering losses were
computed analytically in perturbative regime for two colliding gaussian-shaped wave
packets. Furthermore these wavepackets were assumed to evolve without losses and without
spreading. The number of elastically scattered atoms as a function of time was found in
the form
\begin{eqnarray}
\label{radka}\mathcal{S}(t)&=&\left(\frac{Na}{\sigma}\right)^2\mathrm{Erf}\left(\frac{\sqrt
2\hbar Q}{m\sigma}t\right),
\end{eqnarray}
where $\sigma$ is a width of the gaussian wave-packets and $Q$ is the wave vector
corresponding to the absolute value of the momentum of each of the wave-packets in the
center of mass frame. The same quantity might be calculated approximately under analogous
assumptions using the stochastic classical noise. The equation for $\psi_B$ in this case
reads
\begin{eqnarray}
\label{psi_b}\imath\hbar\partial_t\psi_B&=&-\frac{\hbar^2}{2m}\nabla^2\psi_B
+gN\left(|\psi_B|^2+2|\psi_1|^2+2|\psi_{2}|^2\right)\psi_{B}\nonumber\\
&+&2gN\psi_{1}\psi_{2} (\psi_B+\xi_{12})^*,
\end{eqnarray}
where $\psi_{1,2}$ are two counter-propagating gaussian wavefunctions. The approximations
of \cite{Bach} amounts in retaining on the right hand side of equation \ref{psi_b} only
the last term. The approximate solution obtained this way gives the number of elastically
scattered atoms as a function of time in the form
\begin{eqnarray}
\label{scatt}\mathcal{S}_{stoch}(t)&=&A_{12}\Delta
t\frac{4\pi\hbar}{m(2Q)}\left(\frac{Na}{\sigma}\right)^2\mathrm{Erf}\left(\frac{\sqrt
2\hbar Q}{m\sigma}t\right).
\end{eqnarray}
Comparing (\ref{radka}) with (\ref{scatt}) we obtain $A_{12}=\frac{m(2Q)}{4\pi\hbar\Delta
t}$. As we anticipated $A_{12}$ depends on the relative velocity of wavepackets $\psi_1$
and $\psi_2$, which in our case is equal $(2Q)$. Note that once $A_{ij}$'s are determined
our numerical approach has no more adjustable parameters. In Fig. \ref{fig1} we are
comparing the solution of (\ref{radka}) with a numerical solution of full equation
(\ref{psi_b}). Note the growing discrepancy between perturbative and non-perturbative
results for larger losses. They result from bosonic enhancement present in the
non-perturbative regime.  We also stress that in the non-perturbative regime the
stochastic noise is crucial at the early stage of evolution. It may even be dropped from
equation (\ref{psi_b}) when Bose enhancement takes effect. This is why the strength of
classical noise may be determined in the perturbative regime. Finally, we point out that
some analogies regarding the break down of perturbative approach were found in the study
of atom-molecule conversion within positive-P representation \cite{Olsen,Poulsen}.

\begin{figure}
\centering
\includegraphics[scale=0.35]{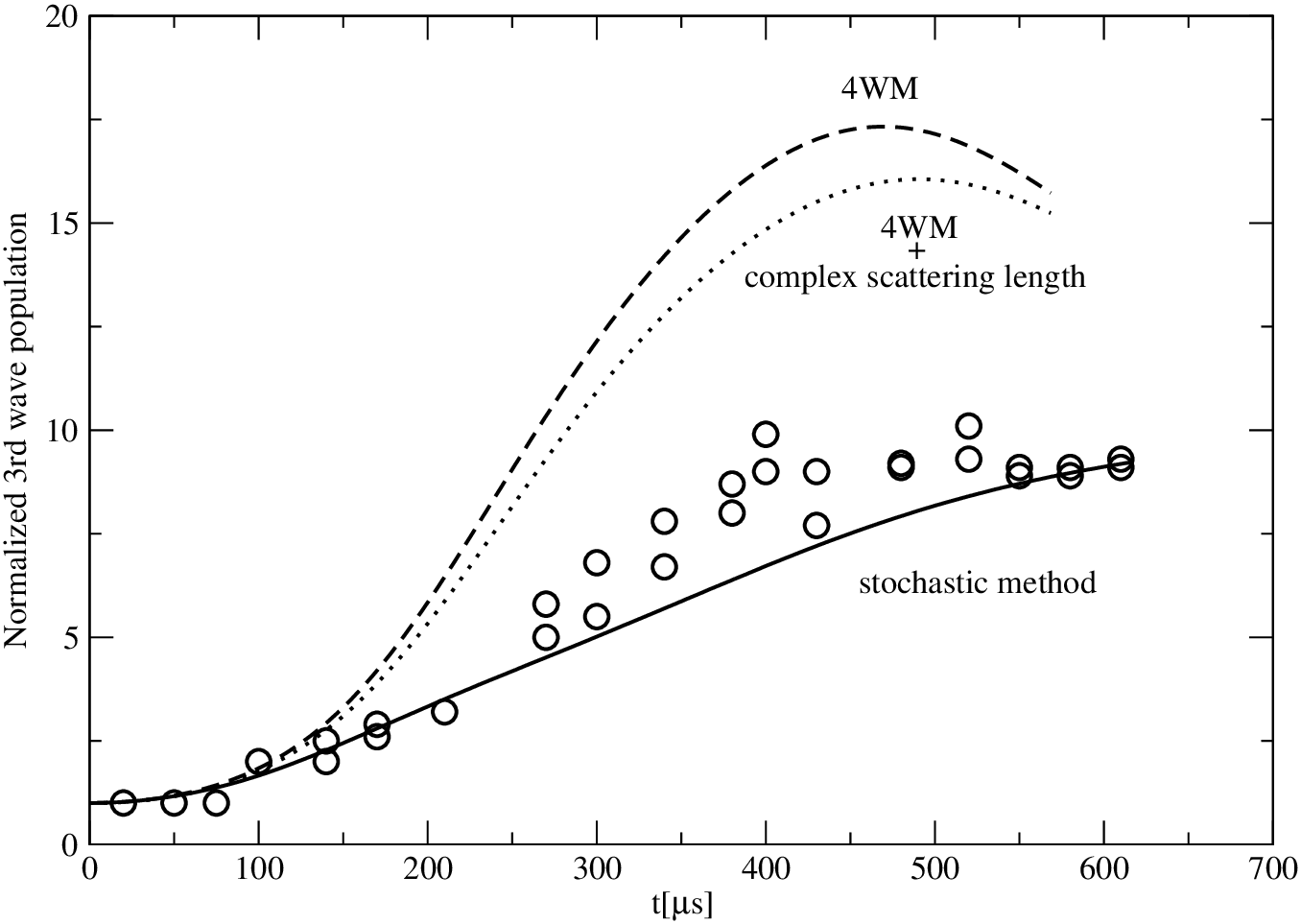}
\caption{Population of the third wavepacket normalized to the initial seed population as
a function of time (continuous line). Parameters used in the simulation correspond to
experiment performed in the Ketterle group ~\cite{Vogels}. Total number of atoms equal to
5mln. Also shown: circles - experimentally measured values, and dashed and dot lines -
solutions of the Gross-Pitaevskii equation with real and complex scattering length
respectively.} \label{fig2}
\end{figure}

With the equations (\ref{4wm_1}-\ref{4wm_2}) fully determined we turn our attention to
the recent experiment from MIT ~\cite{Vogels}. This experiment, due to the large value of
the ratio of collision to nonlinear timescales, had very large number of elastically
scattered atoms. Experimental configuration consists of two initial wavepackets of equal
strength ($\approx N/2$ atoms in each) and the third wavepacket of just a tiny fraction
of N. Magnetic trap used to generate the Sodium condensate had frequencies of $80$, $80$
and $20$ Hz in axial direction, hence it has a shape of a cigar. Applied optical Bragg
pulses to create moving wavepackets propagated approximately at the same angle of
$\approx 0.5$ rad with respect to the long axis of the  condensate corresponding to a
relative velocity of $20$ mm/s. In two series of 4WM measurements chemical potential of
the condensate was $2.2$ and $4.4$ kHz, which we identified as corresponding to $5$ and
$30$ million atoms respectively. In Figure \ref{fig2} we plot the population of the third
wave-packet (the one that was initially seeded) as a function of time. The circles are
the experimental data extracted from paper ~\cite{Vogels}. Several theoretical curves are
plotted. The dashed line represents results neglecting all elastic scattering losses. The
dotted line accounts for the losses by means of the complex scattering length
\cite{Band}. We see that neither of the curves reproduces experimental results. Our
stochastic method gives the solid line which is much closer to the experimental data. It
has been computed with the parameters of the experiment including the initial number of
atoms infered from the paper as being equal to 5mln.

\begin{figure}
\centering
\includegraphics[scale=0.35,angle=0]{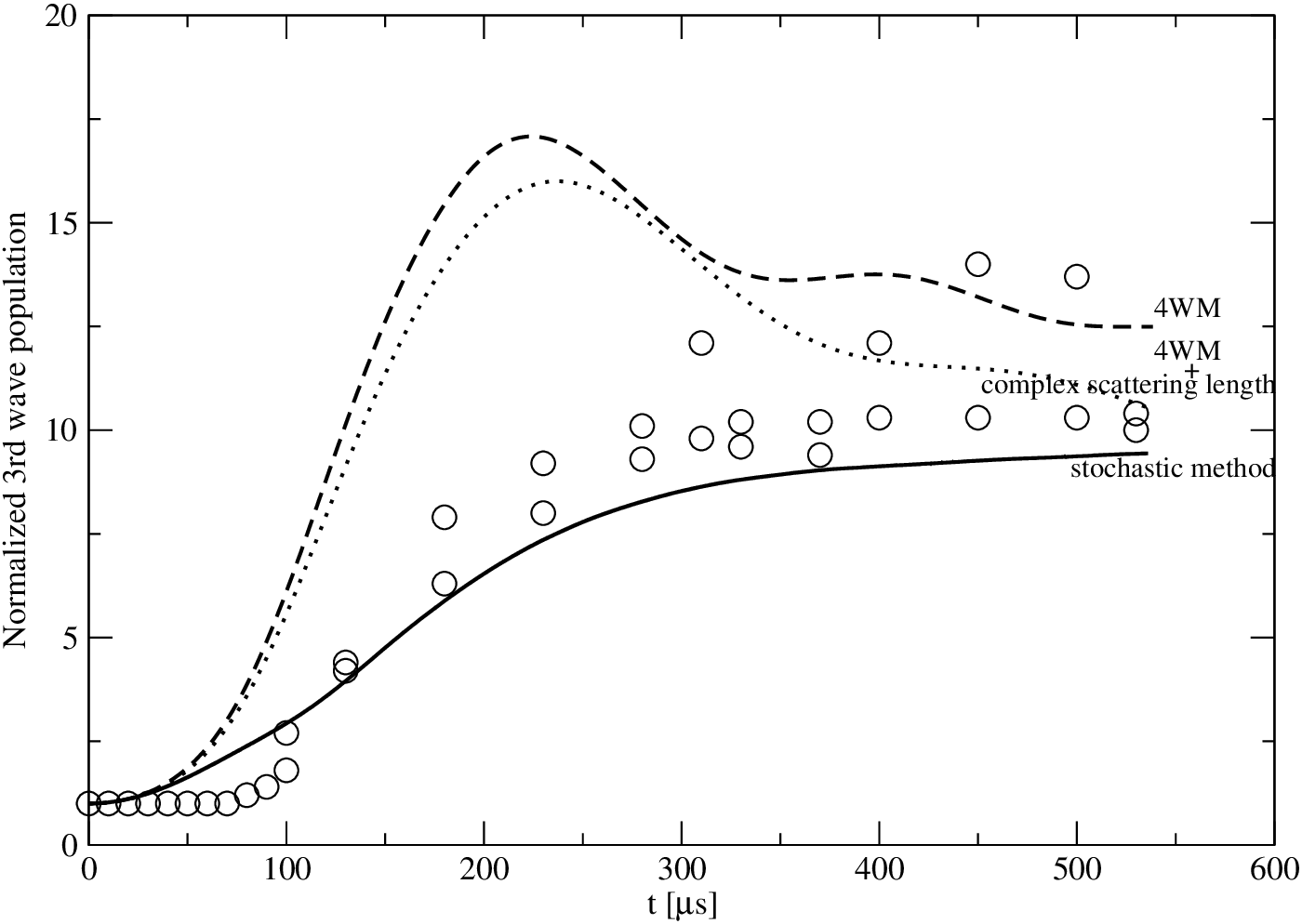}
\caption{Population of the third wavepacket normalized to the
initial seed population as a function of time (continuous line).
Parameters used in the simulation correspond to experiment
performed in the Ketterle group ~\cite{Vogels}. Total number of
atoms equal to 30 mln. Also shown: circles - experimentally
measured values, and dashed and dot lines -  solutions of the
Gross-Pitaevskii equation with real and complex scattering length
respectively.} \label{fig3}
\end{figure}


In Fig.\ref{fig3} similar comparison is made for larger sample of 30mln atoms. Again our
results reproduce the experimental data very well. We feel that remaining discrepancy
(our results seem to be consistently under experimental points) is due to
indistinguishability of BEC and thermal atoms in the region of the momentum space
occupied by BEC.


In conclusion: We have formulated the classical stochastic field
method that accounts for the quantum fluctuations responsible for
spontaneous initiation of various atom optics processes. For
instance we can treat oscillations between atomic and molecular
condensates triggered by optical or magnetic field effects
~\cite{Olsen,Poulsen}. The method is then applied to the atomic
4WM. It gives for the first time excellent agreement with the
recent MIT experiment, where the scattering losses where
substantial.

We acknowledge stimulating discussions with Mariusz Gajda, Piotr Deuar and Keith Burnett.
The authors would like to acknowledge support from KBN Grant 2P03 B4325 (J. Ch.), Polish
Ministry of Scientific Research and Information Technology under grant
PBZ-MIN-008/P03/2003 (M. T., K.R ).

\end{document}